\documentclass[conference,10pt]{IEEEtran}
\IEEEoverridecommandlockouts
\usepackage{cite}
\usepackage{amsmath,amssymb,amsfonts}
\usepackage{algpseudocode} 
\usepackage{algorithm}
\usepackage{graphicx}
\usepackage{textcomp}
\usepackage{xcolor}
\def\BibTeX{{\rm B\kern-.05em{\sc i\kern-.025em b}\kern-.08em
    T\kern-.1667em\lower.7ex\hbox{E}\kern-.125emX}}
\usepackage{cuted}
\usepackage{amsthm}

\theoremstyle{remark}

\usepackage{mathtools}
\usepackage{titlesec}
\usepackage[caption=false,font=footnotesize]{subfig}

\titlespacing{\subsection}{0pt}{0.5\baselineskip}{0.1\baselineskip}

\begin{document}
\title{Online UAV Trajectory Planning Under QoS Constraints to Mobile Users in Urban Environments }

\author{
    \IEEEauthorblockN{Chenrui Qiu\IEEEauthorrefmark{1}, Loizos Kanaris\IEEEauthorrefmark{2}, Yongxu Zhu\IEEEauthorrefmark{3}, Tasos Dagiuklas\IEEEauthorrefmark{1}}
    \IEEEauthorblockA{\IEEEauthorrefmark{1} School of Computer Science and Digital Technologies, London South Bank University, London, UK} 
   \IEEEauthorblockA{\IEEEauthorrefmark{2} Sigint Solutions Ltd, Nicosia, Cyprus}
    \IEEEauthorblockA{\IEEEauthorrefmark{3} School of Information Science and Engineering, Southeast University, Nanjing, China}
    \IEEEauthorblockA{\{qiuc3, tdagiuklas\}@lsbu.ac.uk,  l.kanaris@sigintsolutions.com, yongxu.zhu@seu.edu.cn}
}

\maketitle

\begin{abstract}
This paper studies real-time trajectory planning and radio resource allocation for a single uncrewed aerial vehicle (UAV) serving multiple mobile ground users in an urban environment. The downlink system considers heterogeneous user mobility, where independent users and group users coexist and interact. To ensure reliable communication, quality-of-service (QoS) constraints are imposed by requiring the instantaneous data rate of each user to satisfy a minimum threshold whenever feasible. A capacity limited high-altitude platform (HAP)-assisted wireless fronthaul is further considered to capture practical network-side transmission limitations. Under these constraints, the UAV updates its position at each time slot, while QoS-aware bandwidth and power are jointly allocated under total bandwidth and transmit power constraints to maximize system throughput. Due to user mobility and urban blockages, the resulting problem is highly nonconvex and time-varying. An online reinforcement learning (RL) based approach is adopted for real-time UAV trajectory optimization. Simulation results show that the proposed method satisfies the QoS, fronthaul, and radio resource constraints and achieves a balanced trade-off between throughput and user fairness.
\end{abstract}


%
\IEEEpeerreviewmaketitle
\section{Introduction}
Uncrewed aerial vehicles (UAVs) are increasingly considered as flexible aerial platforms for communication and edge intelligence in beyond 5G/6G networks. By exploiting their controllable 3-$\textit{D}$ mobility, UAVs can rapidly form favorable air-to-ground links, reshape coverage, and provide on-demand services to users in scenarios where terrestrial infrastructure is insufficient or highly dynamic. This capability is particularly attractive for mobile-user environments\cite{hou2022joint}, in which user positions and traffic demands evolve over time, making static network planning ineffective. In such settings, UAV trajectory optimization becomes a core design dimension that directly impacts throughput, latency, coverage continuity, and energy consumption.

UAV trajectory optimization under user mobility has been studied as a sequential decision problem. In \cite{yang2022dynamic}, trajectory control was jointly designed with computation offloading in UAV-enabled MEC systems, highlighting the time-coupled nature of UAV mobility. Building on this view, \cite{qian2022path} applied Monte Carlo Tree Search for online path planning in dynamic UAV-aided wireless systems to improve adaptability under evolving user states.

The feasibility of such online trajectories is constrained by energy consumption. In \cite{becvar2022energy}, the power cost of flying base stations serving mobile users was analyzed, showing that mobility-aware designs must jointly consider propulsion and communication energy. Beyond feasibility, trajectory design has been coupled with service scheduling. Joint user scheduling and trajectory optimization was proposed by \cite{yuan2022joint} to minimize completion time, which was later extended to UAV-relaying-assisted MEC with moving users in \cite{qi2023completion}.
To handle the resulting high-dimensional optimization, learning-based approaches have been introduced. In \cite{song2022evolutionary}, evolutionary multi-objective reinforcement learning was used for joint trajectory control and task offloading, and \cite{wei2024joint} further studied split offloading and trajectory scheduling in UAV-enabled IoT networks. These works embed user mobility implicitly in learned policies.

From a communication-centric perspective, trajectory optimization for moving users has also been studied under structured network models. In \cite{hou2022joint}, joint resource allocation and trajectory design was investigated for multi-UAV systems with moving users using learning-assisted optimization, while \cite{li2023optimal} extended trajectory design to integrated sensing and communications networks with moving users. Physical-layer enhancements were incorporated in \cite{zargari2022user}, where user scheduling and UAV trajectory were jointly optimized in IRS-assisted UAV networks.

Recent studies indicate that user mobility structure plays a critical role in trajectory design. In \cite{yan2023joint}, joint trajectory control and resource allocation was studied for mobile group users, showing that correlated motion significantly affects feasible trajectories. Related observations were reported in \cite{hu2025collaborative}, where collaborative positioning for multiple moving users was investigated in UAV-enabled ISAC systems.

Despite these advances, most existing studies rely on simplified user mobility assumptions, such as fixed group memberships, predictable motion patterns, or static user sets. In realistic urban environments, user mobility is heterogeneous and time-varying, where individual and group-based users may coexist, interact, merge, and split under dynamic blockage conditions. This motivates the study of a more practical scenario with mixed user mobility and time-varying blockages.
The main contributions of this paper are summarized as follows:
i) We investigate an urban UAV-enabled communication system in which heterogeneous mobile users, including independent users and group users, coexist and interact under mixed mobility patterns. The UAV updates its position at each time slot to adapt to time-varying user locations and urban blockages, while being constrained by a capacity-limited HAP-assisted wireless fronthaul.
ii) An online reinforcement learning–based per-slot trajectory planning method has been developed to address the resulting highly nonconvex and time-varying optimization problem under user mobility, urban propagation effects, and UAV kinematic constraints.
iii) We incorporate a QoS-aware joint bandwidth and transmit power allocation scheme under total bandwidth and transmit power constraints, which ensures user QoS requirements while improving overall system throughput.



\section{System Model}
\begin{figure}
\centering
\includegraphics[scale=0.25]{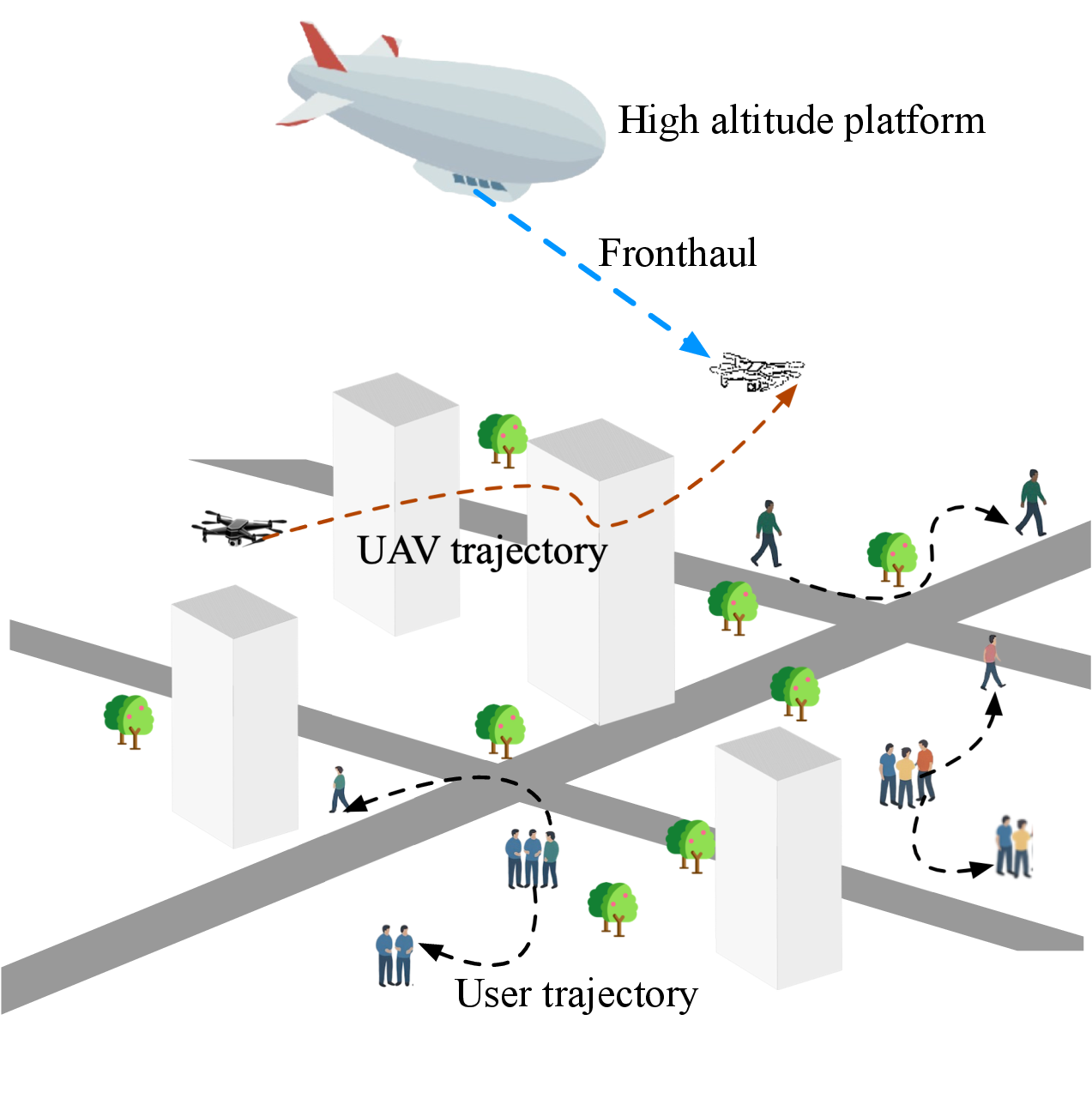}            
\caption{UAV-Assisted network model with individual and group users.}
\label{model}
\end{figure}

We consider the downlink of a HAP-assisted UAV wireless network, as shown in Fig.~\ref{model}, where a single UAV acts as an aerial base station to serve $K$ moving ground users and is connected to the core network via a HAP through a capacity-limited fronthaul link. The users are partitioned into $G$ disjoint groups and a set of individual users. The total communication period $T$ is divided into $N$ equal time slots with duration $\delta$, such that $T = N\delta$. Let $\mathcal{N} \triangleq \{1,2,\ldots,N\}$ denote the set of time-slot indices. Let $\mathbf{w}_k[n] \in \mathbb{R}^2$ denote the horizontal location of ground user $k \in \mathcal{K} \triangleq \{1,\ldots,K\}$ at time slot $n$. The UAV is assumed to fly at a fixed altitude $ H$. Its horizontal position at time slot $n$ is denoted by $\mathbf{q}[n] \in \mathbb{R}^2$, so the UAV trajectory is defined as $\mathbf{Q}=\{\mathbf{q}[1], \cdots,\mathbf{q}[n],\cdots,\mathbf{q}[N]\}$.

\subsection{User Mobility Model} 
\begin{figure}[!htbp]
\centering
\includegraphics[scale = 0.45]{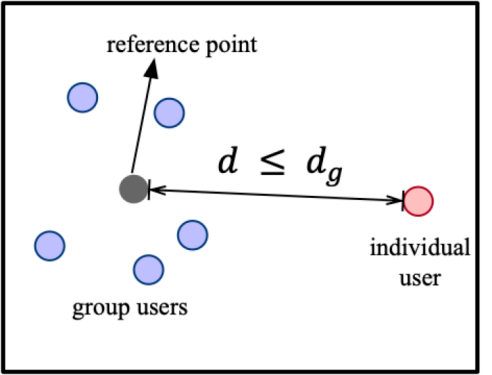}              
\caption{Interaction between RPGM and individual user.}
\label{mobility1}
\end{figure}

To better capture realistic user mobility scenarios, we consider a hybrid mobility model consisting of both group users and individual users, including the interaction between group and individual users. Specifically, one individual could become a member of a group, and one member of a group could becom an individual as Fig. \ref{mobility1}.

\label{mobility}
\vspace{-0.1cm}
We model group users follow reference point group mobility (RPGM), in which the group motion is driven by a reference point and members fluctuate around it. Individual users follow random walk mobility (RWM), moving independently via random steps \cite{6676324}.
Instead of using an abstract displacement vector, we express the RP movement in polar form as
\begin{equation}
\mathbf{w}_r[n] = \mathbf{w}_r[n-1] + v_r[n]
\begin{bmatrix}
\cos\theta_r[n]\\
\sin\theta_r[n]
\end{bmatrix},
\label{eq:rp_update_polar}
\end{equation}
where $v_r[n]\ge 0$ and $\theta_r[n]\in[0,2\pi)$ denote the step length and moving direction of the RP at slot $n$, respectively.
The position of user $k$ in the group is modeled as a bounded perturbation around the RP as
\begin{equation}
\mathbf{w}_k[n] = \mathbf{w}_r[n] + r_k[n]
\begin{bmatrix}
\cos\phi_k[n]\\
\sin\phi_k[n]
\end{bmatrix},
\quad \forall k,
\label{eq:user_update_polar}
\end{equation}
where $r_k[n]\in[0,r_d^{\max}]$ denotes the deviation distance and $\phi_k[n]\in[0,2\pi)$ denotes the deviation direction. 

For individual users, the position of an individual user $k$ evolves as
\begin{equation}
\mathbf{w}_k[n] = \mathbf{w}_k[n-1] + v_k[n]
\begin{bmatrix}
\cos\theta_k[n]\\
\sin\theta_k[n]
\end{bmatrix}, \quad \forall k,
\label{eq:individual_update}
\end{equation}
where $v_k[n]\ge 0$ and $\theta_k[n]\in[0,2\pi)$ denote the step length and moving direction.

An individual user may \emph{join} a nearby group with probability $p_{\mathrm{join}}$, provided that at least one reference point is within an attachment radius $d_{g}$ as shown in Fig. \ref{mobility1}. Specifically, define the candidate set as
\begin{equation}
\mathcal{G}_k[n] \triangleq 
\left\{\, g \,\middle|\, \left\lVert \mathbf{w}_k[n]-\mathbf{w}_g[n]\right\rVert \le d_{g} \right\}.
\end{equation}
If $\mathcal{G}k[n]\neq\emptyset$, then with probability $p_{\mathrm{join}}$ user $k$ switches to group mode and selects a group with the nearest reference point,
Conversely, a group member may \emph{leave} the group with probability $p_{\mathrm{leave}}$ if it drifts beyond a detachment radius $d{g}$.

At each time slot, both the UAV and users perform a velocity-based position update followed by boundary reflection; obstacle avoidance is enforced by rejecting any candidate position or trajectory segment intersecting buildings and reselecting a feasible direction with reduced step length.

\subsection{Propagation Model}
To better capture practical air-to-ground propagation in different environments, we consider both line-of-sight (LoS) and non-line-of-sight (NLoS) links depending on the blockage condition. So we introduce an indicator $\delta_k[n]\in\{0,1\}$, where $\delta_k[n]=1$ denotes an LoS link and $\delta_k[n]=0$ denotes an NLoS link, determined by the environment.

The distance between UAV $\mathbf{q}[n]$ and user $\mathbf{w}_k[n]$  is
$
d_k[n] = \sqrt{\|\mathbf{q}[n]-\mathbf{w}_k[n]\|^2 + H^2}.
$

The large-scale channel power gain is modeled as
$
\beta_k[n] = \frac{\beta_0}{d_k[n]^{\alpha_{\rm{L}/\rm {NL}}}},
$
where $\beta_0$ is the channel power gain at the reference distance 1 m, $\alpha_{\rm{L}/\rm{NL}}$ is the LoS or NLoS path-loss exponent.

To model the small-scale fading coefficient $g_k[n]$ with a scalable baseline suitable for large-scale simulations, Rician fading for LoS links and Rayleigh fading for NLoS links.
\begin{equation}
g_k[n]\sim
\begin{cases}
\mathcal{CN}\!\left(\sqrt{\dfrac{\kappa}{\kappa+1}}e^{j\phi_k[n]},\ \dfrac{1}{\kappa+1}\right), & \delta_k[n]=1,\\[10pt]
\mathcal{CN}(0,1), & \delta_k[n]=0,
\end{cases}
\end{equation}
where $\kappa$ denotes the Rician $K$-factor and $\phi_k[n]\in[0,2\pi)$ represents the phase of the LoS component.
With this normalization, $\mathbb{E}\!\left[|g_k[n]|^2\right]=1$ holds for both LoS and NLoS cases.

Thus, the achievable rate of user k in slot n is denoted as
\vspace{-0.05cm}
\begin{equation} \label{rate}
    {\rm {R}}_{k}[n] = b_k[n] \log_{2}{(1+\frac{p_{k}[n]\beta_k[n]g_k[n]}{b_k[n]N_0})},
\end{equation}
where $b_k[n]$ and $p_{k}[n]$ are the bandwidth allocation and the transmit power from UAV to user $k$ in slot n, $N_0$ is noise power spectrum density.

\subsection{LoS Detection}
Consider a UAV located at $\mathbf{q}=(x_q,y_q,H)$ and a ground user located at $\mathbf{w_k}=(x_k,y_k,0)$.
The 3-$\textit{D}$ LoS path between the UAV and the user is modeled as the straight line segment
\begin{equation}
\mathbf{r}(\lambda)=\mathbf{q}+t(\mathbf{w}-\mathbf{q})=\begin{pmatrix}
x_q+t(x_k-x_q)\\
y_q+t(y_k-y_q)\\
z(t)
\end{pmatrix},\\\quad t\in[0,1],
\end{equation}
whose height component $z(t)=H(1-t)$.
Each building $i$ is modeled as a vertical prism with a rectangular ground footprint and roof height $h_i$.
A building is a potential blocker only if the 2-$\textit{D}$ projection of the line segment $(x(t),y(t))$ intersects the footprint of building $i$,
which yields an overlapping parameter interval $t\in[t_{\mathrm{in}},t_{\mathrm{out}}]$.
Since $z(t)$ is monotonically decreasing in $t$, the minimum height of the LoS path over the building footprint occurs at $t_{\mathrm{out}}$, i.e.,
\begin{equation}
z_{\min}=z(t_{\mathrm{out}})=H\bigl(1-t_{\mathrm{out}}\bigr).
\end{equation}
The UAV-to-user link is declared NLoS if there exists at least one building $i$ such that $z_{\min}\le h_i$,
meaning that the 3-$\textit{D}$ LoS path intersects the building volume; otherwise, if $z_{\min}>h_i$ for all intersecting buildings, the link is classified as LoS.

\subsection{Problem Formulation}
In this mobility user scenario, we consider jointly optimizing the trajectory of UAV. Sepcially, when optimizing the achievable rate for UAV to moving users at each slot, account for the transmit power, fronthaul with limited capacity and safety with blockage. Since the HAP–UAV fronthaul is typically designed to be highly reliable and exhibits much slower variations than the UAV–to-user links, its capacity is assumed to be constant over the considered time horizon.

The instantaneous UAV-to-user links may experience severe attenuation or even NLoS conditions. Under the per-slot total bandwidth and power budgets and the fronthaul capacity constraint, enforcing the instantaneous QoS requirement ${ {R}}_{k}[n] \geq { {R}}_{min}$ for all users and all slots may render the optimization infeasible in certain slots. To capture the “whenever feasible” QoS principle while keeping the per-slot optimization always feasible, we introduce nonnegative slack variables $\leq 0 s_k[n]\leq \frac{R_{min}}{10}$ to quantify unavoidable QoS violations. When the original QoS constraints are feasible, the optimal slack variables become zero automatically; otherwise, the proposed formulation yields the least-violating feasible solution. 
Thus, the optimization problem is formulated as
\vspace{-0.4cm}
\begin{subequations}\label{constrain} \begin{align} \mathbb{P}_1: \max_{\mathbf{Q}, p_k[n], b_k[n]}~ & \sum_{n=1}^{N}\sum_{k=1}^{K} { {R}}_{k}[n], \label{constrain.a} \\ \text{s.t.} \quad & {R}_{k}[n] + s_k[n] \ge {R}_{\min}, \ \forall k,n,\label{constrain.b} \\ & \sum_{k=1}^{K} { {R}}_{k}[n] \leq C_f, \ \forall n, \label{constrain.c} \\ & 0 \leq x_q[n] \leq x_{\max}, \ \forall n,\label{constrain.d} \\ & 0 \leq y_q[n] \leq y_{\max}, \ \forall n,\label{constrain.e} \\ & \left\lVert \mathbf{q}[n]-\mathbf{q}[n-1]\right\rVert \leq {v}_{\max}, \ \forall n,\label{constrain.f} \\ & b_k[n]\ge 0, \sum_{k=1}^{K} b_k[n] \leq B_{max}, \ \forall n,\label{constrain.g} \\ & p_k[n]\ge 0, \sum_{k=1}^{K} p_k[n] \leq P_{max}, \ \forall n,\label{constrain.h} \end{align} \end{subequations}
where \eqref{constrain.b} guarantee the QoS, \eqref{constrain.c} restricts the aggregate user rate by the fronthaul capacity, \eqref{constrain.d} and \eqref{constrain.e} guarantee the position of UAV must be within the map boundary, \eqref{constrain.f} is the limitation of UAV moving speed, \eqref{constrain.g} states that the bandwidth constrains for all time slots, \eqref{constrain.h} indicates the sum of transmit power of UAV to each user for one slot is limited by a maximum transmit power.

\subsection{QoS-Aware Bandwidth--Power Allocation}
\label{bandwidth-power}
We develop a low-complexity QoS-aware bandwidth--power allocation strategy for mobile
users under limited bandwidth, transmit power, and fronthaul capacity.

Let $a_k[n]\triangleq \frac{\beta_k[n]\bar g_k[n]}{N_0}>0$. The achievable downlink rate is
$
R_k[n]=b_k[n]\log_2\!\left(1+\frac{a_k[n]\,p_k[n]}{b_k[n]}\right).
$
For a fixed UAV trajectory, the per-slot resource allocation is
\vspace{-0.2cm}
\begin{subequations}\label{p2}
\begin{align}
\mathbb{P}_2:\ \max_{\{b_k[n],p_k[n],s_k[n]\}} \quad
& \sum_{k=1}^{K} R_k[n] \label{eq:P2_obj}\\
\text{s.t.}\quad \eqref{constrain.b}, \eqref{constrain.c},\eqref{constrain.g},\eqref{constrain.h}
\end{align}
\end{subequations}
Since $b_k[n]\log_2\!\left(1+\frac{a_k[n]\,p_k[n]}{b_k[n]}\right)$ is the perspective of a concave function, $R_k[n]$ is concave in
$(b_k[n],p_k[n])$. Solving \eqref{p2} optimally at each slot is computationally demanding,
thus we adopt the heuristic procedure in Algorithm~\ref{alg1}.

Given $\{b_k[n]\}$, the minimum power to satisfy $R_k[n]\ge \hat R_k[n] = R_{min}-s_k[n]$ is
\begin{equation}\label{pmin}
p_{k,\min}[n]=\frac{\big(2^{\hat R_k[n]/b_k[n]}-1\big)b_k[n]}{a_k[n]}.
\end{equation}
If $\sum_k p_{k,\min}[n]\le P_{\max}$, the QoS targets are feasible; otherwise, bandwidth is
reallocated. We define
\begin{equation}\label{chi}
\chi_k[n]=\frac{a_k[n]}{2^{\hat R_k[n]/b_k[n]}-1},
\end{equation}
where users with smaller $\chi_k[n]$ require higher power to meet the same target rate, and are
prioritized in bandwidth allocation. We update bandwidth via
\vspace{-0.4cm}
\begin{equation}\label{bw_update}
b_k[n]\leftarrow B_{\max}\frac{w_k[n]}{\sum_{j=1}^{K} w_j[n]},\qquad
w_k[n]\triangleq \frac{1}{\chi_k[n]}.
\end{equation}
If feasibility cannot be achieved due to the fronthaul or power limits, the slack variables
$s_k[n]$ are increased for the most demanding users (smallest $\chi_k[n]$) until
$\sum_k \hat R_k[n]\le C_f$ and $\sum_k p_{k,\min}[n]\le P_{\max}$.

With a feasible $(\{b_k[n]\},\{p_{k,\min}[n]\})$, the remaining power
$P_{\mathrm{rem}}=P_{\max}-\sum_k p_{k,\min}[n]$ is allocated by water-filling:
\begin{equation}\label{water}
\Delta p_k^{\star}[n]=\Big[\mu-\frac{b_k[n]}{a_k[n]}\Big]^+,
\end{equation}
where $\mu$ is chosen such that $\sum_k \Delta p_k^{\star}[n]\le P_{\mathrm{rem}}$ and
$\sum_k R_k[n]\le C_f$. The final power is $p_k[n]=p_{k,\min}[n]+\Delta p_k^{\star}[n]$,
which ensures $R_k[n]\ge R_{\min}$ whenever feasible; otherwise, $R_k[n]\ge R_{\min}-s_k[n]$.


\begin{algorithm}[t]
\caption{QoS-aware heuristic bandwidth-power allocation for $\mathbb{P}_2$}
\label{alg1}
\begin{algorithmic}[1]
\Require $\mathbf{q}[n]$, $\{a_k[n]\}$, $(B_{\max},P_{\max},R_{\min},C_f)$, $I_{\max}$, \qquad step: $(\Delta b,\Delta s)$.
\Ensure $\{b_k[n]\}$, $\{p_k[n]\}$, $\{s_k[n]\}$.
\State Initialize $b_k[n]\gets B_{\max}/K,\ s_k[n]\gets 0,\ \forall k$.
\For{$i=1$ to $I_{\max}$}
    \State Set $\hat R_k[n]\gets R_{\min}-s_k[n],\ \forall k$.
    \While{$\sum_k \hat R_k[n] > C_f$}
        \State $k^\star\gets \arg\min_k \chi_k[n]$; $s_{k^\star}[n]\gets s_{k^\star}[n]+\Delta s$; update $\hat R_k[n]$.
    \EndWhile
    \State Compute $p_{k,\min}[n]$ (replace $R_{\min}$ by $\hat R_k[n]$ in \eqref{pmin}).
    \If{$\sum_k p_{k,\min}[n]\le P_{\max}$} \textbf{break} \EndIf
    \State Compute $\chi_k[n]$ (replace $R_{\min}$ by $\hat R_k[n]$ in \eqref{chi}).
    \State $k_1\gets \arg\min_k \chi_k[n],\ k_2\gets \arg\max_k \chi_k[n]$.
    \State $b_{k_1}[n]\gets b_{k_1}[n]+\Delta b,\ b_{k_2}[n]\gets \max\{b_{k_2}[n]-\Delta b,0\}$; normalize $\sum_k b_k[n]=B_{\max}$.
\EndFor
\While{$\sum_k p_{k,\min}[n] > P_{\max}$}
    \State $k^\star\gets \arg\min_k \chi_k[n]$; $s_{k^\star}[n]\gets s_{k^\star}[n]+\Delta s$; update $\hat r_k[n],\ p_{k,\min}[n],\ \chi_k[n]$.
\EndWhile
\State Set $p_k[n]\gets p_{k,\min}[n]$, $P_{\rm rem}\gets P_{\max}-\sum_k p_{k,\min}[n]$.
\If{$P_{\rm rem}>0$}
    \State Obtain $\Delta p_k^{\star}[n]=\Big[\mu-\frac{b_k[n]}{a_k[n]}\Big]^+$ by bisection on $\mu$
    \Statex \hspace{1.2em} s.t. $\sum_k \Delta p_k^{\star}[n]\le P_{\rm rem}$ and $\sum_k R_k[n]\le C_f$.
    \State $p_k[n]\gets p_{k,\min}[n]+\Delta p_k^{\star}[n],\ \forall k$.
\EndIf
\State \Return $\{b_k[n]\},\{p_k[n]\},\{s_k[n]\}$.
\end{algorithmic}
\end{algorithm}

\section{UAV Trajectory Optimization}
\label{sec:traj}
Problem $\mathbb{P}_1$ couples the UAV trajectory $\mathbf{Q}$ with the per-slot bandwidth--power allocation $\{b_k[n],p_k[n]\}$ through the achievable rates $\{R_k[n]\}$. Owing to user mobility, blockage-induced LoS/NLoS transitions, and the nonconvex kinematic constraints, solving $\mathbb{P}_1$ via conventional optimization across all slots is computationally complex for real-time operation. Motivated by this, we adopt a reinforcement learning (RL) framework to learn an online trajectory control policy that maps the observed system state to a feasible movement action at each slot as Algorithm. \ref{alg:rl_traj_with_alg1}.
\vspace{-0.1cm}
\subsection{MDP Modeling}
\label{subsec:mdp}
We model the UAV trajectory control as a Markov decision process (MDP) \cite{9872114} over a finite horizon of $N$ slots. At slot $n$, the agent observes state $\mathbf{s}[n]$, selects action $\mathbf{a}[n]$, the environment evolves to $\mathbf{s}[n+1]$, and the agent receives an immediate reward $r[n]$. The policy is denoted by $\pi(\mathbf{a}[n]|\mathbf{s}[n])$.
(1) State:
The state is constructed to capture the position of the UAV and users.
\vspace{-0.3cm}
\begin{equation} 
\mathbf{s}[n] \triangleq
\Big[
{x_q}[n],~{y_q}[n],~
{x_k}[n],~{y_k}[n]~,{k\in\mathcal{K}}
\Big],
\label{eq:state_def}
\end{equation}
(2) Action:
At slot $n$, the action $\mathbf{a}_n$ represents the displacement of the UAV from $\mathbf{q}_n$ to $\mathbf{q}_{n+1}$, defined as
$
\mathbf{a}_n = \mathbf{q}_{n+1} - \mathbf{q}_n,
$
where $\|\mathbf{a}[n]\| \leq v_{max} \delta$.
(3) Reward:
Given $\mathbf{q}[n]$ and user locations, the achievable rates $\{R_k[n]\}$ are computed by solving $\mathbb{P}_2$ using the QoS-aware heuristic bandwidth--power allocation described in Section~\ref{bandwidth-power}. To promote proportional fairness while discouraging QoS violations and overly aggressive maneuvers, we define the per-slot reward as
\vspace{-0.3cm}
\begin{align}
r[n] & \triangleq
\sum_{k=1}^K \log\!\big(\varepsilon + R_k[n]\big)
- \lambda \sum_{k=1}^K \big[R_{\min}+s_k[n]-R_k[n]\big]^+ \nonumber\\
&- \eta \bigg[\sum_{k=1}^{K} R_k[n] - C_f\bigg]^+
- \mu \|\mathbf{a}[n]\|^2,
\label{eq:reward}
\end{align}
where $\log\!\big(\varepsilon + R_k[n]\big)$ enforces proportional fairness among users, prioritizing rate improvements for low-throughput users and avoiding over-reliance on high-throughput users for reward gains.  $\varepsilon>0$ is a small constant, ensures numerical stability, precluding the logarithm from approaching negative infinity when $\{R_k[n]\}$ is near zero and unstable RL policy updates. $[\cdot]^+=\max\{\cdot,0\}$, and
$\lambda$, $\eta$, and $\mu$ are positive weighting coefficients.

\subsection{Learning Method and Online Execution}
\label{subsec:learning}

The MDP defined in \eqref{eq:state_def}--\eqref{eq:reward} involves continuous
state and action spaces. We adopt a policy-gradient reinforcement learning
approach with an actor--critic architecture to learn the UAV trajectory control
policy $\pi(\mathbf{a}|\mathbf{s})$.

Specifically, proximal policy optimization (PPO) is employed due to its stable training behavior in continuous control problems. During training, each episode consists of $N$ slots with randomized user initial locations and movements are generated according to the chosen mobility model. At each slot, the agent executes the action $\mathbf{a}[n]$ and receives the reward in \eqref{eq:reward} after solving the per-slot resource allocation problem.

During online execution, the trained actor network generates the UAV action
$\mathbf{a}[n]$ via a single forward pass at each slot. The resulting online
complexity is dominated by policy inference and per-slot resource allocation,
which scales linearly with the number of users.

\begin{algorithm}[t]
\caption{RL-based UAV trajectory optimization with QoS-aware resource allocation}
\label{alg:rl_traj_with_alg1}
\begin{algorithmic}[1]
\Require Time slots $N$, training iteration $M$, map boundary $(x_{\max},y_{\max})$, mobility constraint $v_{\max}$, system parameters $(B_{\max},P_{\max},R_{\min},C_f)$
\State Initialize actor-critic networks and PPO parameters
\For{$m=1$ to $M$}
    \State Randomly initialize user positions and mobility states; initialize UAV position $\mathbf{q}[1]$ by K-means
    \For{$n=1$ to $N$}
        \State Observe state $\mathbf{s}[n]$
        \State Sample action $\mathbf{a}[n]$, enforce $\|\mathbf{a}[n]\|\le v_{\max}\delta$
        \State {Algorithm~\ref{alg1}} to obtain $\{b_k[n],p_k[n]\}$ or {Infeasible}
        \If{Infeasible}
            \State Set reward $r[n]\leftarrow -\infty$ and terminate rollout
            \State \textbf{break}
        \Else
            \State $R_k[n]$, $\forall k$;
            \State Reward $r[n]$ using \eqref{eq:reward};
            \State Store transition $(\mathbf{s}[n],\mathbf{a}[n],r[n],\mathbf{s}[n+1])$
        \EndIf
    \EndFor
    \State Update PPO policy and value networks using collected trajectories
\EndFor
\State \Return trained policy $\pi_\theta(\mathbf{a}|\mathbf{s})$.
\end{algorithmic}
\end{algorithm}

\vspace{-0.1cm}
\section{Simulation and Results}
In this section, simulation results are presented to evaluate the performance of the proposed UAV trajectory optimization scheme in a dynamic urban environment.
We consider a downlink UAV communication system over a 2-$\textit{D}$ urban area
$\mathcal{A}=[0,300]\times[0,300]~\mathrm{m}^2$.
In the simulations, the users consist of 3 groups with $4$ users and $10$ individual users,
yielding $K=22$ users in total.
All users are mobile with a maximum speed $r_d^{\max}=2~\mathrm{m/s}$.
The group radius is set to $d_g=20~\mathrm{m}$.
The UAV is subject to a maximum horizontal speed constraint of $v_{\max}=16~\mathrm{m/s}$ and operates at a fixed altitude of $H=100~\mathrm{m}$.
The total simulation duration is $120~\mathrm{s}$ with a time resolution of $1~\mathrm{s}$.
Users may dynamically join or leave groups according to human-like mobility behaviors, with the joining and leaving probabilities set to $p_{\mathrm{join}}=p_{\mathrm{leave}}=0.5$.
The noise power spectral density is set to $N_0=-170~\mathrm{dBm/Hz}$.
The minimum rate requirement is $R_{\min}=1~\mathrm{Mbps/Hz}$, the maximum transmit power is $P_{\max}=2~\mathrm{W}$, the maximum available bandwidth is $B_{\max}=20~\mathrm{MHz}$, and the fronthaul limitation set as $500~\mathrm{Mbps}$, the channel power gain at the reference distance 1 m is $\beta_0=-50~\mathrm{dB}$.

\begin{figure}[!t]
\begin{minipage}[t]{0.495\linewidth}
    \centering
    \includegraphics[width=4.4cm, height=4cm]{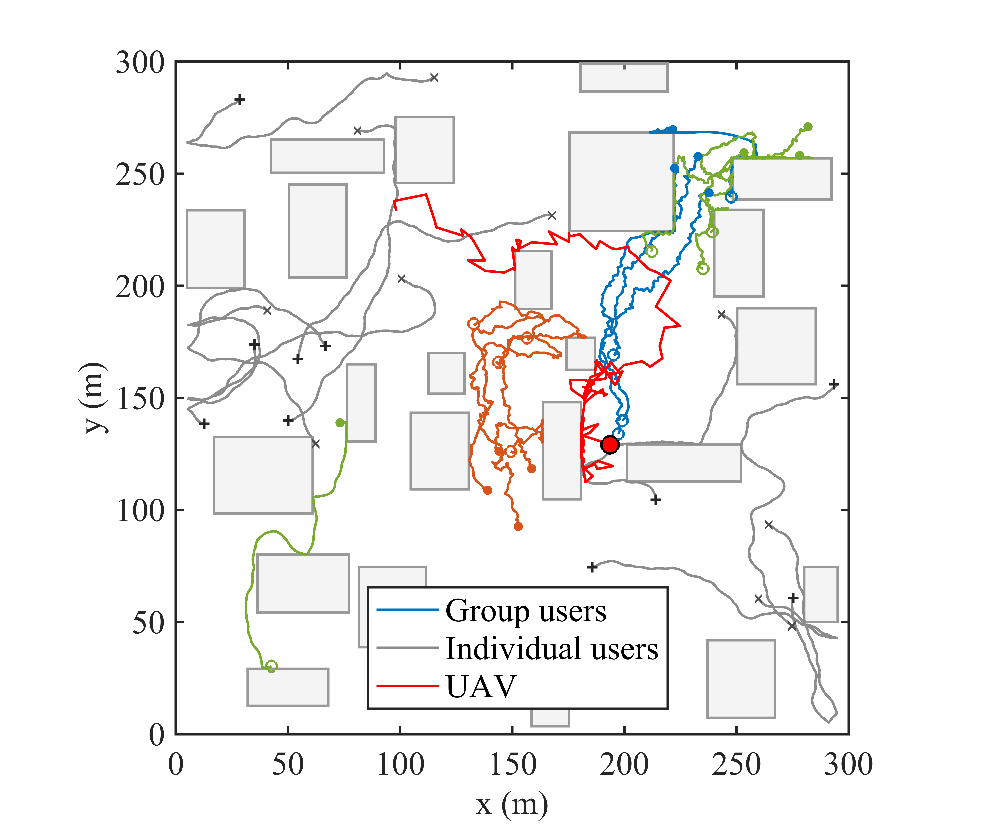}
\caption{Trajectories of users and UAV.} \label{fig3}
\end{minipage}%
\hfill
\begin{minipage}[t]{0.495\linewidth}
    \centering
    \includegraphics[width=4.3cm, height=4cm]{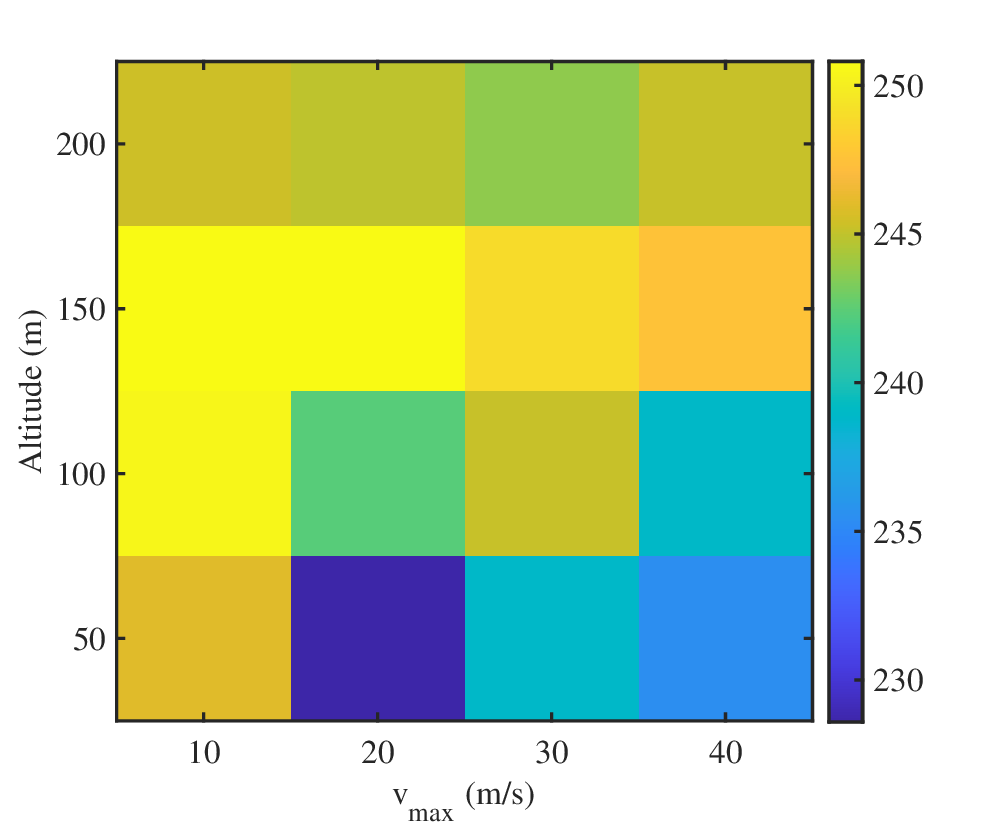}
\caption{Average throughput versus UAV maximum speed  
under different flight altitude.} \label{fig4}
\end{minipage}
\end{figure}
Fig. \ref{fig3} shows the trajectories of the UAV and mobile users at 120s. It can be observed that users exhibit heterogeneous mobility behaviors, including both individual and group movements. Despite such complex user dynamics and the presence of obstacles, the UAV trajectory adapts to the overall user distribution and maintains close proximity to active user clusters. This indicates that the UAV is capable of effectively tracking user mobility and dynamically adjusting its flight path to serve both individual users and groups.

Fig.~\ref{fig4} shows the average throughput over a 120s horizon versus the UAV altitude $H$ and maximum velocity $v_{\max}$. The throughput exhibits a clear 2-$\textit{D}$ and non-monotonic dependence on these flight parameters. At low $H$, building blockage increases NLoS links and limits throughput; as $H$ increases, improved LoS enhances throughput, while excessive $H$ increases path loss and degrades performance. Increasing $v_{\max}$ enables faster repositioning to mitigate unfavorable channels, but the gain saturates due to diminishing returns. Hence, the best performance is achieved at an intermediate region ($H=150$~m and \mbox{$v_{\max}=20$, $30$~m/s} in this scenario). \emph{The optimal $H$ and $v_{\max}$ depend on the terrain and building density/height distribution.}

\begin{figure}
\centering
\subfloat[][\label{fig5a}]{\includegraphics[width=4.3cm, height=4cm]{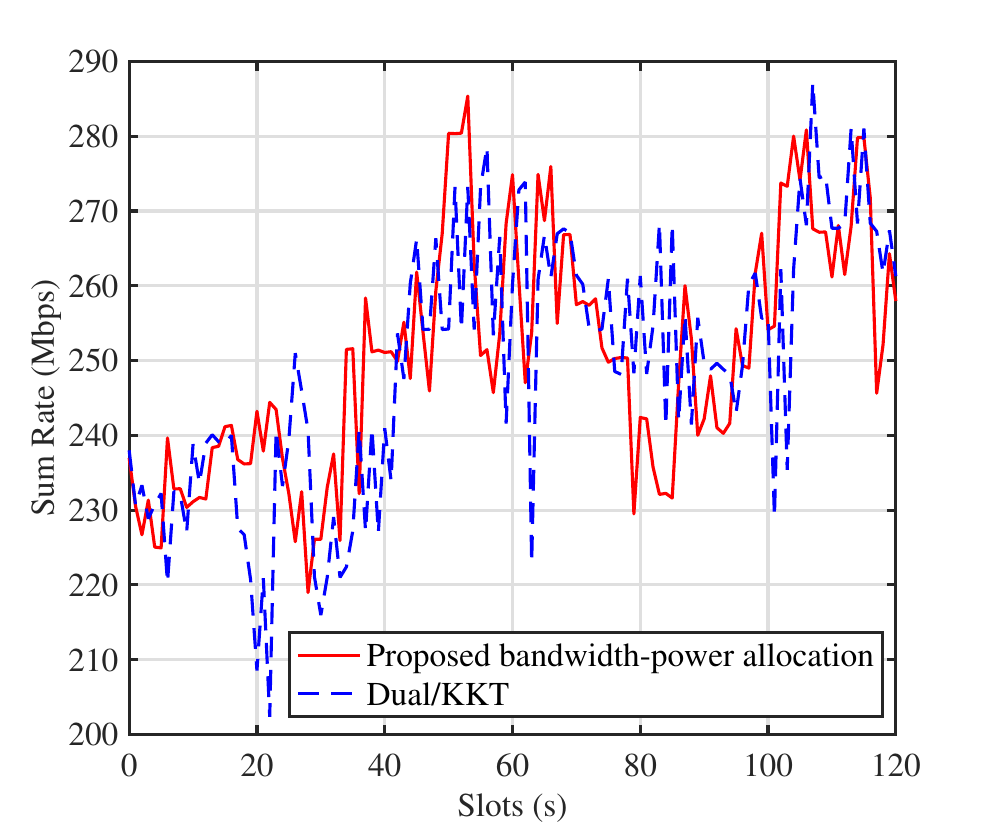}}
\hfill
\subfloat[][\label{fig5b}]{\includegraphics[width=4.3cm, height=4cm]{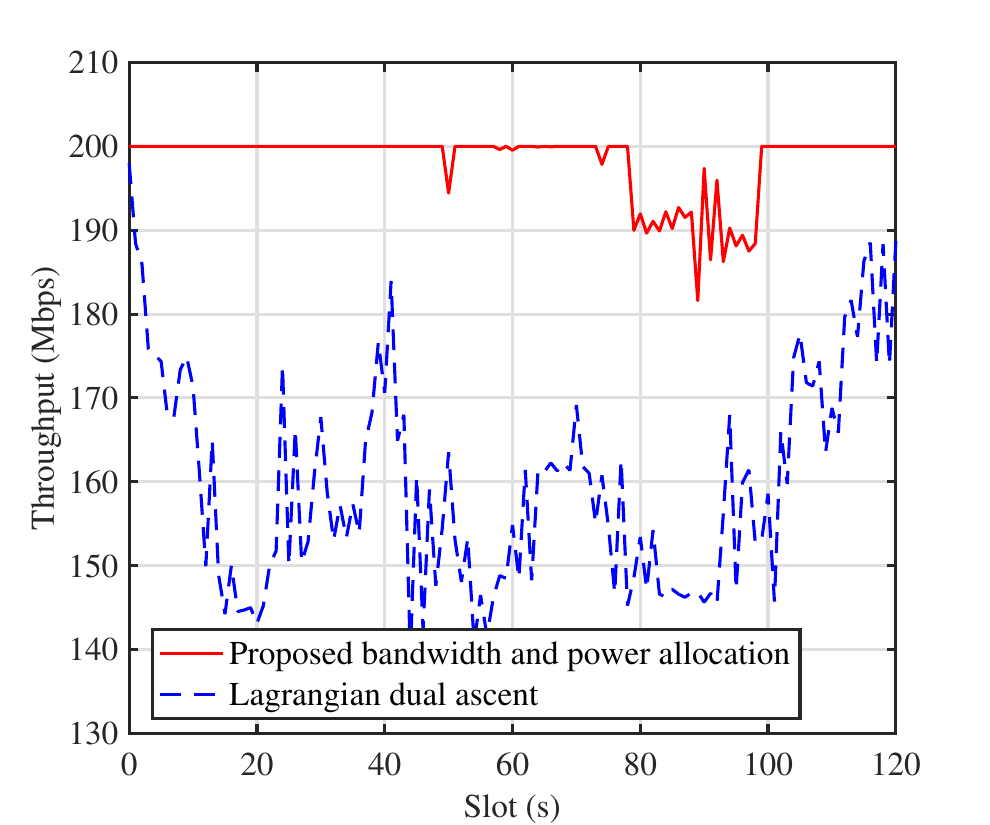}}
\caption{Time evolution of the system sum rate achieved by the proposed bandwidth--power allocation and Lagrangian dual ascent under different fronthaul capacity limitation (a) $C_f = 500 \rm{Mbps}$ and (b) $C_f = 200 \rm{Mbps}$.}
\label{fig5}
\end{figure}
Fig. \ref{fig5} compares the throughput performance between proposed QoS-aware heuristic bandwidth–power allocation algorithm and the Lagrangian dual ascent\cite{10176326}. It can be observed that when in high front capacity the proposed method closely follows the optimal benchmark across the entire time horizon. Despite minor performance degradation at certain time instants, the proposed algorithm achieves near-optimal performance with significantly reduced computational complexity, which is desirable for practical implementations. However, when low fronthaul capacity, Lagrangian dual ascent solution has bad performance.

\begin{figure}
\begin{minipage}[t]{0.495\linewidth}
    \centering
    \includegraphics[width=4.3cm, height=4cm]{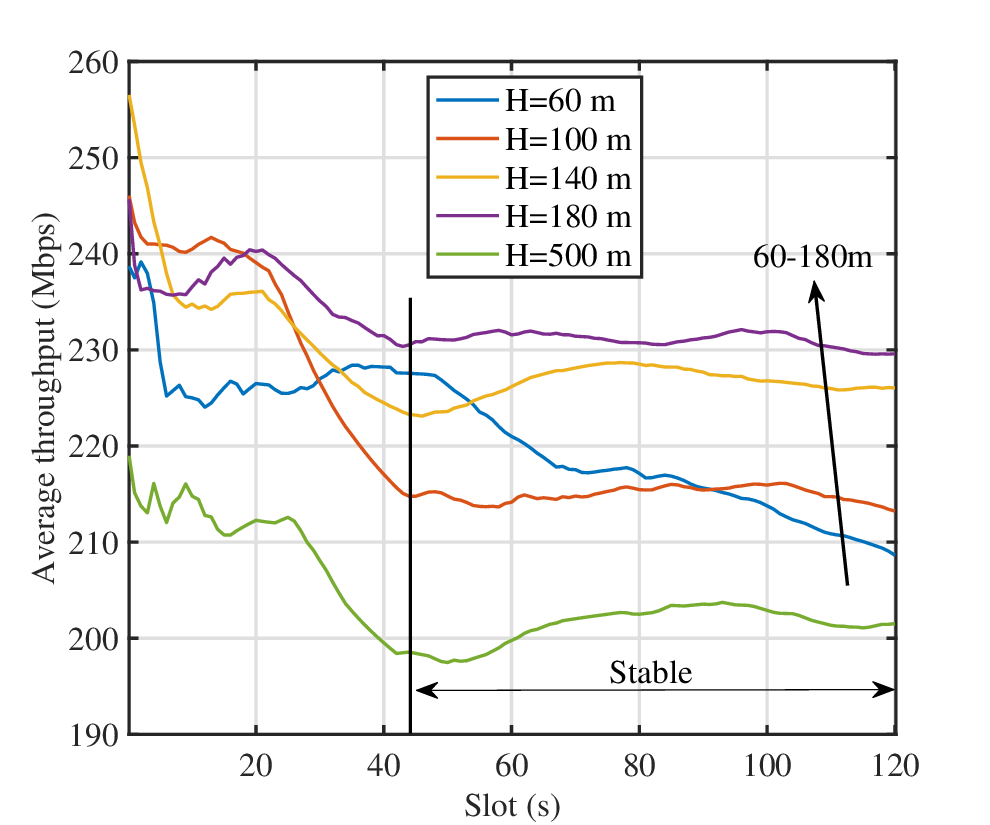}
\caption{Average throughput versus time for different altitudes of UAV, with $K = 20 $.} \label{1}
\end{minipage}%
\hfill
\begin{minipage}[t]{0.495\linewidth}
    \centering
    \includegraphics[width=4.3cm, height=4cm]{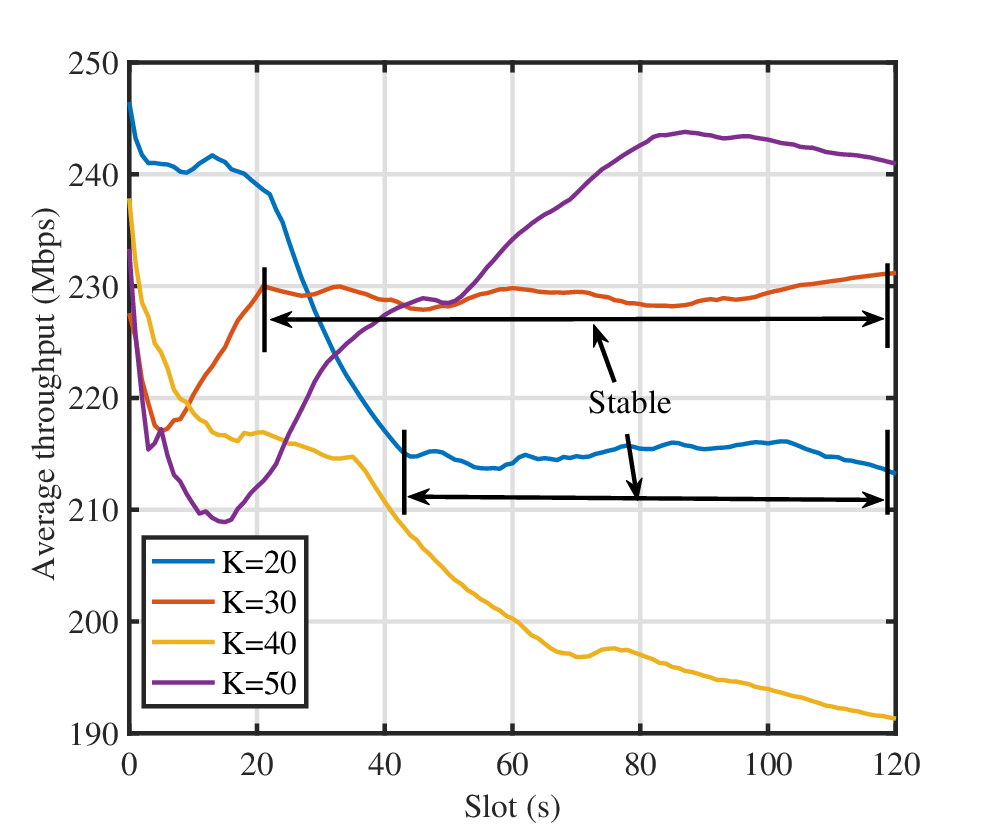}
\caption{Average throughput versus time for different numbers of users K, with $ H = 100 \rm m$.} \label{2}
\end{minipage}
\end{figure}
It is observed from Fig.~\ref{1} that, although different UAV altitudes exhibit distinct transient behaviors in the initial stage, the throughput curves corresponding to high altitudes show smaller fluctuations after convergence. This indicates that operating at higher altitudes leads to more stable long-term throughput performance, mainly due to the more LoS conditions. While for cxcessive height, the increased path loss caused by the longer link distance dominates the LoS gain, resulting in a reduced average throughput.
From Fig.~\ref{2}, it can be seen that a smaller number of users results in smoother throughput evolution over time. In particular, the cases with $K=20$ and $K=30$ exhibit noticeably reduced temporal variations compared with larger $K$, indicating improved stability in the steady-state regime when the system load is moderate. Overall, the results suggest that higher UAV altitudes and smaller numbers of users lead to reduced temporal fluctuations of the average throughput after convergence, thereby yielding more stable long-term system performance.

\section{Conclusion}
This paper investigated a UAV-enabled communication system serving mobile users in obstacle-dense urban environments. A real-time UAV trajectory planning framework was developed by jointly modeling user mobility and a height-aware 3-$\textit{D}$ LoS/NLoS channel model. Soft QoS requirements and capacity-limited HAP-assisted fronthaul constraints were explicitly incorporated into the problem formulation. Moreover, proportional fairness among users was achieved through joint bandwidth allocation and power control, and an RL based approach was employed to address the resulting nonconvex and time-varying trajectory optimization problem. Simulation results demonstrate that the proposed framework effectively adapts to dynamic user mobility and urban blockages, strictly satisfies QoS and fronthaul constraints, and attains a favorable trade-off between system throughput and user fairness. This work considers a single-UAV scenario. As an important future research direction, multi-UAV cooperative deployment and energy-aware learning-based trajectory optimization merit further investigation.

\section*{Acknowledgment}
This work is financed by the European Commission under the Horizon Europe MSCA programme (HORIZON-MSCA-2024-SE-01-01), Grant Agreement No. 101236523 (AeroNet project).

\bibliographystyle{IEEEtran}
\bibliography{mybib}

@STRING{IEEE_J_COM        = "{IEEE} Trans. Commun."}

@STRING{IEEE_J_WCOM       = "{IEEE} Trans. Wireless Commun."}

@STRING{IEEE_J_VT         = "{IEEE} Trans. Veh. Technol."}

@STRING{IEEE_J_ITS        = "{IEEE} Trans. Intell. Transp. Syst."}

@STRING{IEEE_J_CE        = "{IEEE} Trans. Consum. Electron."}

@STRING{IEEE_TMC        = "{IEEE} Trans. Mob. Comput."}

@STRING{IEEE_TNSE        = "{IEEE} Trans. Network Sci. Eng."}

@ARTICLE{10176326,
  author={Yan, Xuezhen and Fang, Xuming and Deng, Cailian and Wang, Xianbin},
  journal=IEEE_J_WCOM, 
  title={Joint Optimization of Resource Allocation and Trajectory Control for Mobile Group Users in Fixed-Wing UAV-Enabled Wireless Network}, 
  year={2024},
  volume={23},
  number={2},
  pages={1608-1621},
  doi={10.1109/TWC.2023.3290748}}

@INPROCEEDINGS{6676324,
  author={Geng, Feng and Xue, Shengjun},
  booktitle={2013 22nd Wireless and Optical Communication Conference}, 
  title={A comparative study of mobility models in the performance evaluation of MCL}, 
  year={2013},
  volume={},
  number={},
  pages={288-292},
  doi={10.1109/WOCC.2013.6676324}}

@article{hou2022joint,
  title={Joint resource allocation and trajectory design for multi-UAV systems with moving users: Pointer network and unfolding},
  author={Hou, Qiushuo and Cai, Yunlong and Hu, Qiyu and Lee, Mengyuan and Yu, Guanding},
  journal=IEEE_J_WCOM,
  volume={22},
  number={5},
  pages={3310--3323},
  year={2022},
  publisher={IEEE}
}

@article{yang2022dynamic,
  title={Dynamic offloading and trajectory control for UAV-enabled mobile edge computing system with energy harvesting devices},
  author={Yang, Zheyuan and Bi, Suzhi and Zhang, Ying-Jun Angela},
  journal=IEEE_J_WCOM,
  volume={21},
  number={12},
  pages={10515--10528},
  year={2022},
  publisher={IEEE}
}

@article{qian2022path,
  title={Path planning for the dynamic uav-aided wireless systems using monte carlo tree search},
  author={Qian, Yuwen and Sheng, Kexin and Ma, Chuan and Li, Jun and Ding, Ming and Hassan, Mahbub},
  journal=IEEE_J_VT,
  volume={71},
  number={6},
  pages={6716--6721},
  year={2022},
  publisher={IEEE}
}

@article{becvar2022energy,
  title={On energy consumption of airship-based flying base stations serving mobile users},
  author={Becvar, Zdenek and Nikooroo, Mohammadsaleh and Mach, Pavel},
  journal=IEEE_J_COM,
  volume={70},
  number={10},
  pages={7006--7022},
  year={2022},
  publisher={IEEE}
}

@article{yuan2022joint,
  title={Joint user scheduling and UAV trajectory design on completion time minimization for UAV-aided data collection},
  author={Yuan, Xiaopeng and Hu, Yulin and Zhang, Jian and Schmeink, Anke},
  journal=IEEE_J_WCOM,
  volume={22},
  number={6},
  pages={3884--3898},
  year={2022},
  publisher={IEEE}
}

@article{qi2023completion,
  title={Completion time optimization in UAV-relaying-assisted MEC networks with moving users},
  author={Qi, Qiuyi and Shi, Tuo and Qin, Ke and Luo, Guangchun},
  journal=IEEE_J_CE,
  volume={70},
  number={1},
  pages={1246--1258},
  year={2023},
  publisher={IEEE}
}

@article{song2022evolutionary,
  title={Evolutionary multi-objective reinforcement learning based trajectory control and task offloading in UAV-assisted mobile edge computing},
  author={Song, Fuhong and Xing, Huanlai and Wang, Xinhan and Luo, Shouxi and Dai, Penglin and Xiao, Zhiwen and Zhao, Bowen},
  journal=IEEE_TMC,
  volume={22},
  number={12},
  pages={7387--7405},
  year={2022},
  publisher={IEEE}
}

@article{wei2024joint,
  title={Joint Split Offloading and Trajectory Scheduling for UAV-enabled Mobile Edge Computing in IoT Network},
  author={Wei, Yunkai and Wan, Zikang and Xiao, Yinan and Leng, Supeng and Wang, Kezhi and Yang, Kun},
  journal=IEEE_TNSE,
  year={2024},
  publisher={IEEE}
}

@article{li2023optimal,
  title={Optimal UAV trajectory design for moving users in integrated sensing and communications networks},
  author={Li, Yiming and Yuan, Xiaopeng and Hu, Yulin and Yang, Junan and Schmeink, Anke},
  journal=IEEE_J_ITS,
  volume={24},
  number={12},
  pages={15113--15130},
  year={2023},
  publisher={IEEE}
}

@article{zargari2022user,
  title={User scheduling and trajectory optimization for energy-efficient IRS-UAV networks with SWIPT},
  author={Zargari, Shayan and Hakimi, Azar and Tellambura, Chintha and Herath, S},
  journal=IEEE_J_VT,
  volume={72},
  number={2},
  pages={1815--1830},
  year={2022},
  publisher={IEEE}
}

@article{yan2023joint,
  title={Joint optimization of resource allocation and trajectory control for mobile group users in fixed-wing UAV-enabled wireless network},
  author={Yan, Xuezhen and Fang, Xuming and Deng, Cailian and Wang, Xianbin},
  journal=IEEE_J_WCOM,
  volume={23},
  number={2},
  pages={1608--1621},
  year={2023},
  publisher={IEEE}
}

@article{hu2025collaborative,
  title={Collaborative positioning optimization for multiple moving users in UAV-enabled ISAC},
  author={Hu, Yunbo and Zhuo, Xiaoxiao and Meng, Zhenyu and Wu, Wen and Lu, Wenkai and Tang, Liang and Qu, Fengzhong and Bu, Zhiyong},
  journal={IEEE Trans. Cognit. Commun. Networking},
  year={2025},
  publisher={IEEE}
}

@ARTICLE{9872114,
  author={Singh, Rahul and Gupta, Abhishek and Shroff, Ness B.},
  journal={IEEE Trans. Control Network Syst.}, 
  title={Learning in Constrained Markov Decision Processes}, 
  year={2023},
  volume={10},
  number={1},
  pages={441-453},
  doi={10.1109/TCNS.2022.3203361}}
\end{document}